\documentclass[twocolumn,showpacs,showkeys,reprint,amsmath,amssymb,aps,prl]{revtex4-1}

\usepackage{graphicx}
\usepackage{dcolumn}
\usepackage{bm}
\usepackage{tabularx}
\usepackage{amsmath}
\usepackage{xcolor}

\emergencystretch=\maxdimen
\begin{document}

\title{Pressure-induced double-dome superconductivity in kagome metal CsTi$_3$Bi$_5$}

\author{J. Y. Nie,$^{1,*}$ X. F. Yang,$^{1,*,\#}$ X. Zhang,$^{1,*,\S}$ X. Q. Liu,$^{2,*}$ W. Xia,$^{2,3}$
	D. Z. Dai,$^1$ C. C. Zhao,$^1$ C. P. Tu,$^{1}$ X. M. Kong,$^1$ X. B. Jin,$^1$ Y. F. Guo,$^{2,3,\dag}$ and S. Y. Li$^{1,4,5,\ddag}$}

\affiliation
{$^1$State Key Laboratory of Surface Physics, Department of Physics, Fudan University, Shanghai 200438, China\\
$^2$School of Physical Science and Technology, ShanghaiTech University, Shanghai 201210, China\\
$^3$ShanghaiTech Laboratory for Topological Physics, Shanghai 201210, China\\
$^4$Collaborative Innovation Center of Advanced Microstructures, Nanjing 210093, China\\
$^5$Shanghai Research Center for Quantum Sciences, Shanghai 201315, China
}

\date{\today}

\begin{abstract}
We present high-pressure resistance measurements up to 40 GPa on recently discovered titanium-based kagome metal CsTi$_3$Bi$_5$. At ambient pressure, CsTi$_3$Bi$_5$ shows no evidence of superconductivity in resistivity and specific heat. By applying pressure, superconductivity emerges and the superconducting transition temperature ${\it T}_{\rm c}$ reaches its first maximum of 1.2 K at $\sim$5 GPa. Then the ${\it T}_{\rm c}$ is suppressed by pressure and cannot be detected around 10 GPa, manifesting as a superconducting dome. Remarkably, upon further increasing pressure above $\sim$13 GPa, another superconducting dome shows up, with the maximum ${\it T}_{\rm c}$ of 0.6 K and ending pressure at $\sim$36 GPa. The variation of ${\it T}_{\rm c}$ displays a clear double-dome shape in the superconducting phase diagram. Our work demonstrates the similarity between CsTi$_3$Bi$_5$ and CsV$_3$Sb$_5$, providing valuable insights into the rich physics of these novel kagome metals.

\end{abstract}


\maketitle



The kagome lattice, a two-dimensional network with corner-sharing triangles, provokes a wide variety of exploration on the correlated electronic phenomena and non-trivial topological states \cite{yin2022topological}. Due to its geometric frustration, kagome lattice naturally possesses unique electronic structures including flat bands, Dirac points, and van Hove singularities, which subsequently give rise to a series of novel quantum phenomena, such as quantum spin liquid \cite{Norman2016colloqioum,yan2011spinliquid}, fractional quantum Hall states \cite{tang2011hightemperature,Neupert2011fractional,bergholtz2015topology}, density waves \cite{Yu2012chiral,wang2013competing}, and unconventional superconductivity \cite{kiesel2012sublattice,kiesel2013unconventional,ko2009doped}. Within this realm, vanadium-based kagome superconductors $\it{A}$V$_3$Sb$_5$ ($\it{A}$ = Cs, Rb and K) have aroused tremendous research interest due to the presence of multiple exotic orders and electronic states, including superconductivity \cite{Ortiz2019new,ortiz2020csv3sb5,jiang2023kagome}, anomalous Hall effect \cite{yang2020giant,yu2021concurrence}, charge density wave (CDW) \cite{jiang2021unconventional,zhao2021cascade,tan2021charge,neupert2022charge}, pair density wave \cite{chen2021roton}, and electronic nematicity \cite{Nie2022charge,Xiang2021twofold}. These states arise within the parent charge density wave phase that appears to simultaneously break both the translational and the rotational symmetry of the lattice. However, the coexisting instabilities in both electron and lattice degrees of freedom make it challenging to identify the intricate interplay between multiple electronic orders.

The titanium-based kagome metal CsTi$_3$Bi$_5$, which is isostructural with CsV$_3$Sb$_5$, has been synthesized more recently \cite{yang2022titaniumbased,Werhahn2022the,yang2022superconductivity}. Compared with CsV$_3$Sb$_5$, the van Hove singularities of CsTi$_3$Bi$_5$ are pushed far above the Fermi level, resulting in the absence of the CDW order \cite{liu2022tunable,wang2023flat,chen2023electrical}. The electronic nematicity of CsTi$_3$Bi$_5$ is dominated by both in-plane and out-of-plane $\it{d}$ orbitals of Ti, as evidenced by the scanning tunneling microscopy (STM) measurements  \cite{yang2022superconductivity,li2022electronic}. Moreover, non-trivial $\mathbb{Z}_2$ topological band structures, including flat bands, Dirac nodal lines, and possible topological surface states, were demonstrated by angle-resolved photoemission spectroscopy (ARPES) and density-functional theory (DFT) calculations \cite{yang2022observation,liu2022tunable,wang2023flat}. Although the exsitence of superconductivity remains controversial due to the conflicting results \cite{yang2022superconductivity,Werhahn2022the,wang2023flat}, CsTi$_3$Bi$_5$ indubitably provides a promising playground for further exploration on the intrinsic electronic structure of the kagome lattice without the influence of the translation symmetry breaking CDW state.

High pressure is a clean and effective method to tune the electronic and structural properties without introducing any impurities. In the case of CsV$_3$Sb$_5$, the CDW order is quickly destabilized under pressure and vanished at 2 GPa, while the superconducting state shows a double-dome evolution with an additional M-shaped modulation in the low-pressure regime \cite{zhao2022nodal,zhang2021pressure, Chen2021highly,yu2022pressure,zhu2021double,yu2021unusual,chen2021double,zheng2022emergent}. The maximum ${\it T}_{\rm c}$ at $\sim$2 GPa in the M-shaped regime coincides with the pressure where CDW state is completely suppressed. Therefore, the origin of the first superconducting dome is attributed to the intricate interplay between superconductivity and the CDW order \cite{yu2021unusual,chen2021double,zheng2022emergent}. Raman measurements and phonon spectra calculations demonstrated that the second superconducting dome should relate to the variation of electronic structure and enhanced electron-phonon coupling due to partial phonon softening \cite{Chen2021highly,yu2022pressure}.

\begin{figure}[b]
	\includegraphics[width=8.6cm]{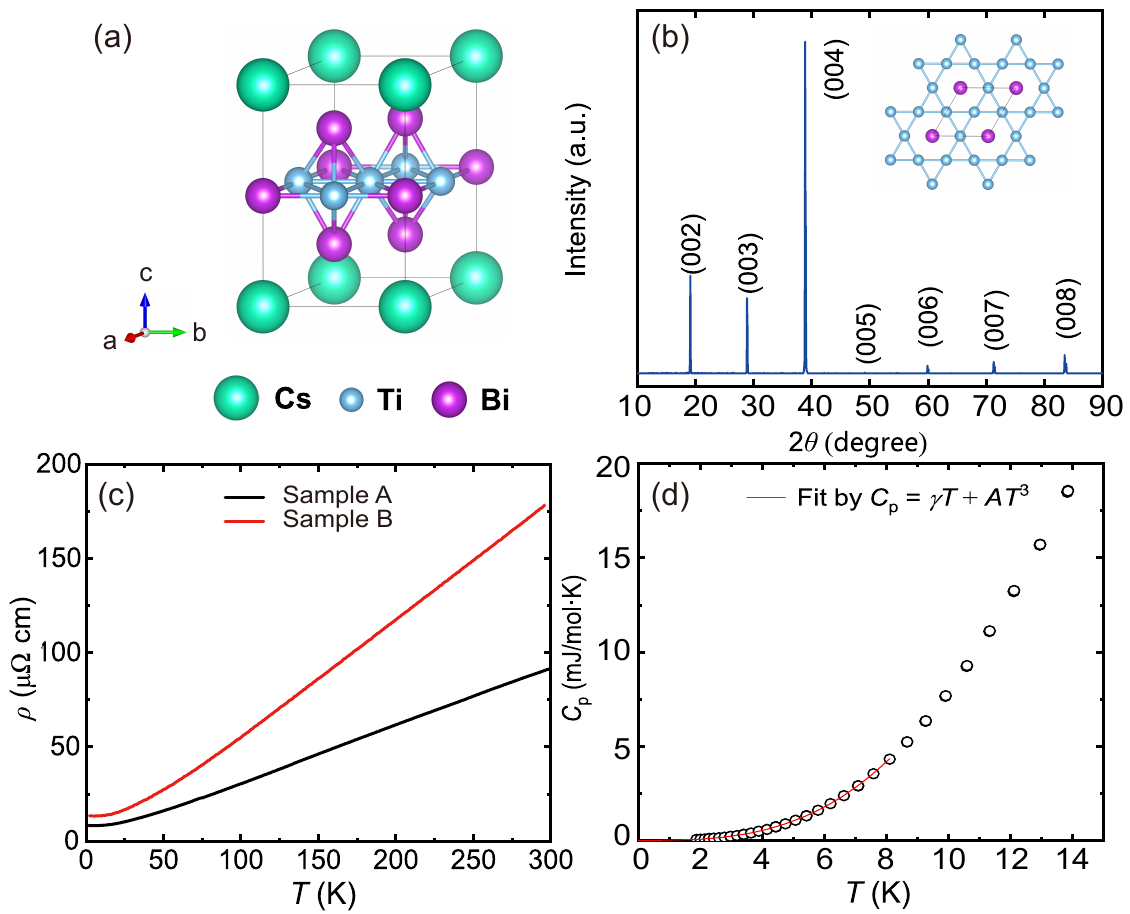}
	\caption{(a) Crystal structure of CsTi$_3$Bi$_5$. The Cs, Ti, and Bi atoms are presented as green, blue, and violet balls. (b) X-ray diffraction pattern of the largest natural surface for CsTi$_3$Bi$_5$ single crystal. The inset of (b) shows kagome net of titanium atoms with bismuth atoms in the hexagons. (c) Temperature dependence of resistivity for CsTi$_3$Bi$_5$ single crystals at ambient pressure. (d) Low-temperature specific heat of CsTi$_3$Bi$_5$ single crystal. The red line shows the fitting of the data blow 8 K to $\it{C}_{\rm p}= \gamma \it{T}+A\it{T}^{\rm 3}$.}
	\label{fig.1}
\end{figure}

\begin{figure}
	\includegraphics[width=8.6cm]{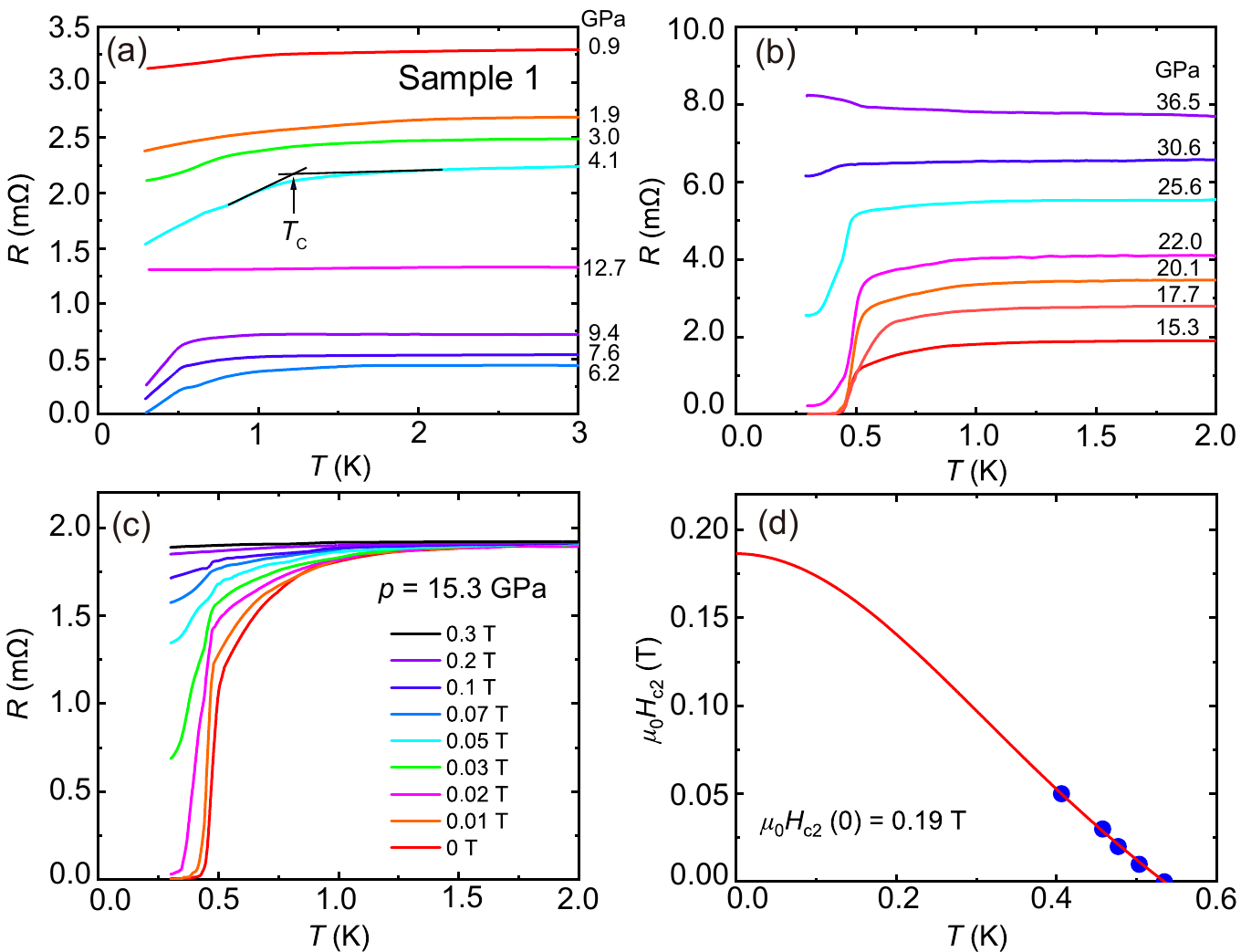}
	\caption{(a) and (b) Temperature dependence of resistance for CsTi$_3$Bi$_5$ Sample 1 under various pressures up to 36.5 GPa. (c) Temperature dependence of resistance for Sample 1 under different magnetic fields at 15.3 GPa. The magnetic field gradually suppresses the superconducting transition. (d) Temperature dependence of the upper critical field $\mu_0{\it H}_{\rm c2}$ at 15.3 GPa. The red line is the fitting of the data to the Ginzburg-Landau formula.}
	\label{fig.2}
\end{figure}

In this Letter, we perform high-pressure resistance measurements on CsTi$_3$Bi$_5$ up to 40 GPa. A double-dome superconductivity is revealed in the temperature-pressure phase diagram, which is reminiscent of that observed in CsV$_3$Sb$_5$. We discuss the underlying physics of the two superconducting domes and point out that the second superconducting dome should also relate to the variation of electronic structure.

Single crystals of CsTi$_3$Bi$_5$ were grown by the self-flux method \cite{Werhahn2022the}. Since CsTi$_3$Bi$_5$ single crystals are sensitive to air, all samples were prepared under an argon atmosphere in order to avoid degradation. For each measurement, we used a fresh piece of CsTi$_3$Bi$_5$ single crystal and peeled off the surface by Scotch tape to ensure the purity. The x-ray diffraction (XRD) measurement was performed using an x-ray diffractometer (D8 Advance, Bruker) with Cu $\it{K}$$\alpha$ radiation (wavelength $\lambda$ = 1.5418 $\text{\AA}$). The resistance and specific heat measurements under ambient pressure were performed using a physical property measurement system (PPMS, Quantum Design), with a very short exposure time to air (several minutes).

High pressure was generated by a diamond anvil cell (DAC) made from Be-Cu alloy. The diamond culet was 400 $\mu$m in diameter. A mixture of epoxy and fine cubic boron nitride (cBN) powder was used to cover the Be-Cu gasket. A 150 $\mu$m sample chamber was drilled at the center of the pre-indented pit, in which a piece of CsTi$_3$Bi$_5$ single crystal was loaded (under an argon atmosphere) with NaCl powder surrounding it as pressure transmitting medium. Platinum foils with a thickness of 4 $\mu$m were used as electrodes. The pressure inside the DAC was determined by monitoring the ruby fluorescence at room temperature each time before and after the measurement. High-pressure resistance was measured in a $^3$He cryostat with the van der Pauw method.

As plotted in Fig. 1(a), CsTi$_3$Bi$_5$ shows a layered hexagonal structure ($\it{P}$6/$\it{mmm}$, No. 191) with lattice constants $a$ = $b$ = 5.787 $\text{\AA}$ and $c$ = 9.206 $\text{\AA}$ \cite{Werhahn2022the}. The Ti-Bi layers and Cs layers are alternately stacked along the $c$ axis. Each Ti-Bi layer contains a kagome net of titanium atoms with bismuth atoms lying in the hexagons as well as above and below the triangles. Figure 1(b) shows the XRD pattern of the largest natural surface for CsTi$_3$Bi$_5$ single crystal. Only $(00l)$ Bragg peaks can be observed, indicating that it is the $ab$ plane. The lattice constant $c$ is determined to be 9.263 $\text{\AA}$. Typical resistivity curves of CsTi$_3$Bi$_5$ single crystal at ambient pressure are plotted in Fig. 1(c). Both sample A and B show metallic behavior and no superconducting transition down to 1.8 K. In Fig. 1(d), the specific heat of CsTi$_3$Bi$_5$ also shows no superconducting transition and the fitting of the data to $\it{C}_{\rm p}= \gamma \it{T}+A\it{T}^{\rm 3}$ below 8 K gives the Sommerfeld coefficient $\gamma$ = 6.9 mJ mol$^{-1}$ K$^{-2}$. However, a tiny superconducting transition with onset temperature of 4.8 K is observed in the magnetic susceptibility of the CsTi$_3$Bi$_5$ single crystal (data not shown), and the superconducting volume fraction is estimated to be approximately 0.005$\%$ at 2 K under the magnetic field of 10 Oe. It very likely comes from a tiny impurity of superconducting CsBi$_2$ with $T_c$ = 4.65 K \cite{Gutowska2022Superconductivity}. Therefore, CsTi$_3$Bi$_5$ itself is a nonsuperconducting kagome metal, as concluded in previous report \cite{Werhahn2022the,chen2023electrical,wang2023flat}.

High-pressure resistance measurements were conducted on two distinct CsTi$_3$Bi$_5$ samples to verify the reproducibility. The samples are labeled as Sample 1 and Sample 2. Figures 2(a) and 2(b) present the low-temperature resistance of Sample 1 in the pressure range from 0.9 to 36.5 GPa. ${\it T}_{\rm c}$ is defined as the temperature where the extrapolation of the normal-state part and the extrapolation of the steep transition part of the resistance curve intersects, as shown in Fig. 2(a) for the 4.1 GPa data. At 0.9 GPa, no superconducting transition is observed for Sample 1 down to 0.3 K, which further confirms the absence of superconductivity at ambient pressure. By increasing pressure, the superconductivity emerges, and ${\it T}_{\rm c}$ is gradually enhanced to 1.21 K at 4.1 GPa. As pressure further increases, ${\it T}_{\rm c}$ shifts to lower temperatures and cannot be detected at 12.7 GPa, showing a superconducting dome. At higher pressure, superconducting transition appears again and ${\it T}_{\rm c}$ increases to a second maximum of 0.63 K at 17.7 GPa. Then ${\it T}_{\rm c}$ decreases again and the superconducting transition disappears completely at 36.5 GPa, showing a second superconducting dome.

\begin{figure}
	\includegraphics[width=8.6cm]{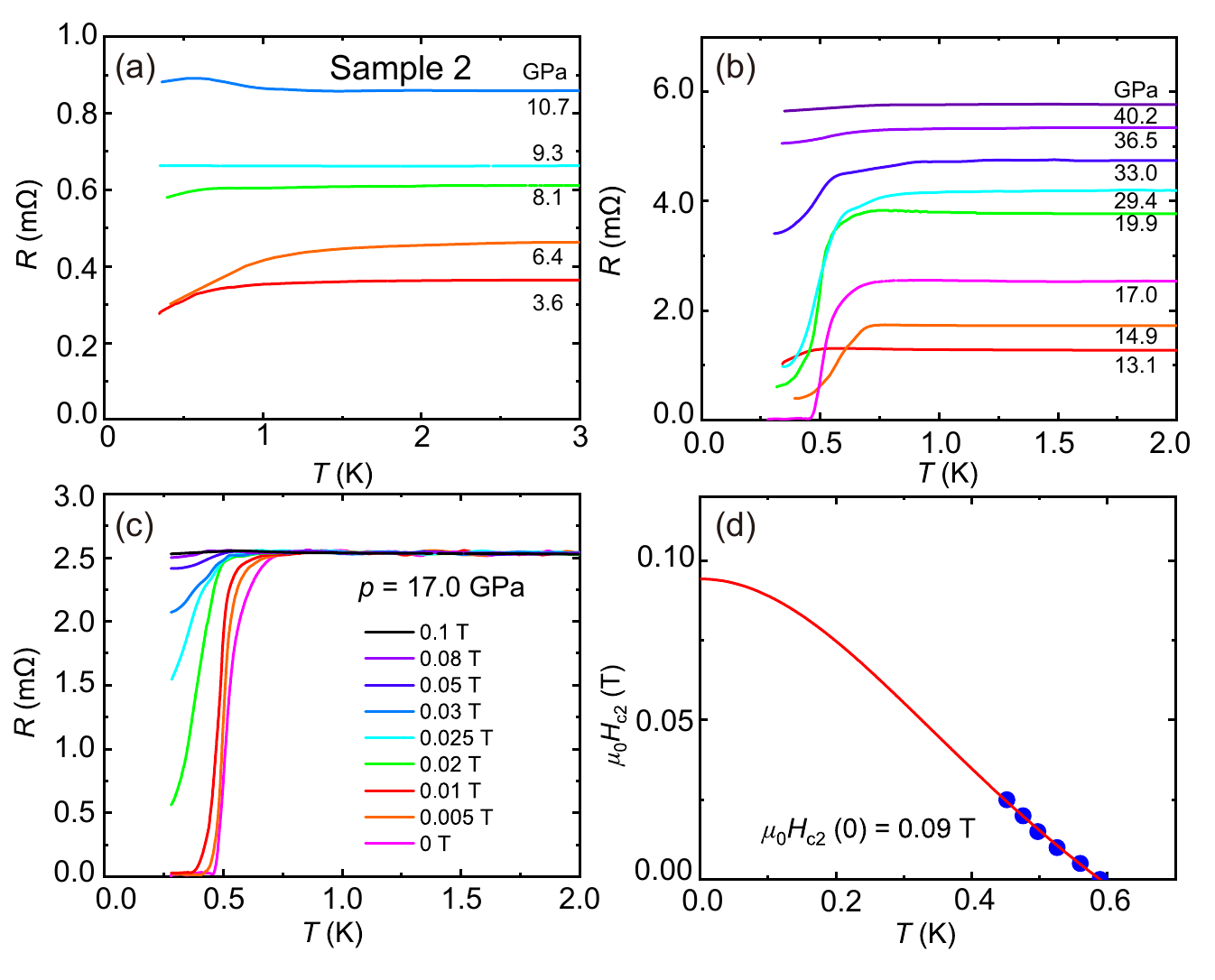}
	\caption{(a) and (b) Temperature dependence of resistance for CsTi$_3$Bi$_5$ Sample 2 under various pressures up to 40.2 GPa. (c) Temperature dependence of resistance for Sample 2 under different magnetic fields at 17.0 GPa. (d) Temperature dependence of the upper critical field $\mu_0\it{H}\rm_{c2}$. The red line is the fitting of the data to the Ginzburg-Landau formula.}
	\label{fig.3}
\end{figure}

To confirm that the resistance drop of CsTi$_3$Bi$_5$ under pressure is superconducting transition, various magnetic fields are applied at 15.3 GPa, as shown in Fig. 2(c). One can see that the resistance drop is monotonically suppressed by increasing the magnetic field until it completely vanishes in 0.3 T. This result demonstrates that the resistance drop under pressure is indeed a superconducting transition. The temperature dependence of upper critical field $\mu_0{\it H}_{\rm c2}$ at 15.3 GPa is summarized in Fig. 2(d). The data can be well fitted by the empirical Ginzburg-Landau (GL) formula $\mu_0{\it H}_{\rm c2}(\it{T}) = \mu_{\rm 0}{\it H}_{\rm c2}(\rm 0)(1-(\it{T}/{\it T}_{\rm c})^{\rm 2})/(\rm 1+(\it{T}/{\it T}_{\rm c})^{\rm 2})$. The $\mu_0{\it H}_{\rm c2}(0)$ is determined to be 0.19 T at 15.3 GPa, much lower than the Pauli paramagnetic limit field ${\it H}_{\rm p}(0)$ = 1.84$\it{T}_{\rm c}\approx$ 0.98 T, indicating the absence of Pauli pair breaking.

\begin{figure}
	\includegraphics[width=8.6cm]{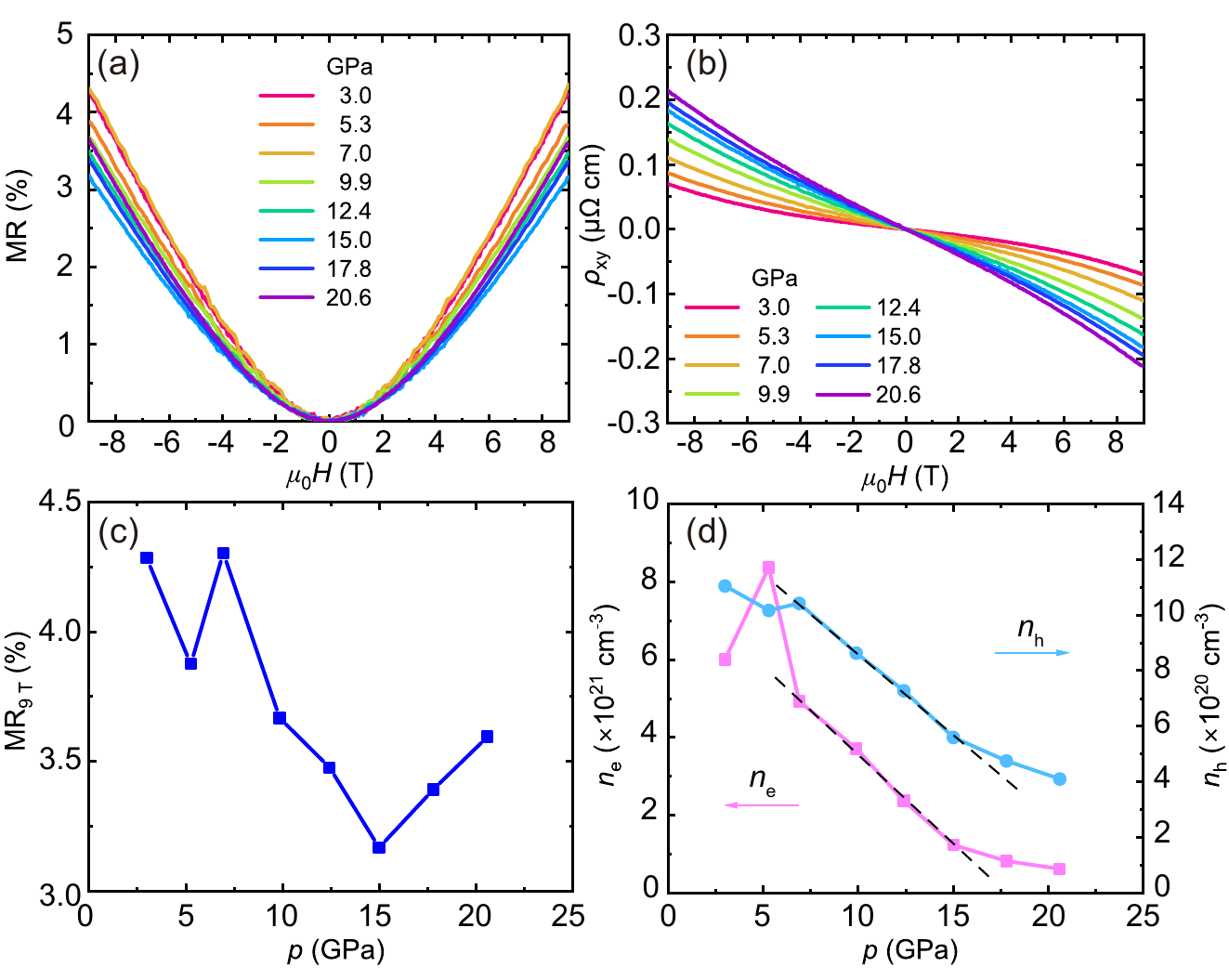}
	\caption{(a) Magnetoresistance (MR) and (b) Hall resistivity of CsTi$_3$Bi$_5$ measured at 10 K and under various pressures in magnetic field applied along the $c$ axis. (c) Pressure dependence of magnetoresistance at 10 K and in 9 T. (d) Pressure dependence of electron carrier density (red) and hole carrier density (blue) obtained from the two-band model. }
	\label{fig.4}
\end{figure}

Similarly, the low-temperature resistance of CsTi$_3$Bi$_5$ Sample 2 in the pressure range from 3.6 to 40.2 GPa is plotted in Figs. 3(a) and 3(b). The superconductivity emerges under pressure, and the ${\it T}_{\rm c}$ of Sample 2 increases to a maximum of 1.17 K at 6.4 GPa. With increasing pressure, the resistance drop gradually moves towards lower temperatures until it is undetectable at 9.3 GPa, showing the first superconducting dome. Upon further compression, superconductivity reemerges with ${\it T}_{\rm c}$ increasing to a second maximum of 0.66 K at 14.9 GPa, followed by a slow reduction. Finally, the superconducting transition vanishes completely at 36.5 GPa, showing the second superconducting dome. The magnetic fields are applied to Sample 2 at 17.0 GPa, as seen in Fig. 3(c). The resistance drop is completely suppressed at 0.1 T. In Fig. 3(d), the $\mu_0\it{H}\rm_{c2}(\it T)$ can also be well fitted by the GL formula, giving $\mu_0\it{H}\rm_{c2}(0)$ = 0.09 T at 17.0 GPa.

Figures 4(a) and 4(b) present the magnetoresistance (MR) and Hall resistivity ($\it{\rho}_{xy}$) of CsTi$_3$Bi$_5$ at 10 K under various pressures up to 20.6 GPa. As pressure increases, the magnitude of magnetoresistance MR = [$\it{\rho}_{xx}(H)$-$\it{\rho}_{xx}(\rm 0)$]/$\it{\rho}_{xx}(\rm 0)$$\times$100\% displays a non-monotonic evolution characterized by a sudden decrease at 5.3 GPa and a reverse increase beyond 15.0 GPa. The pressure dependence of MR at 9 T is plotted in Fig. 4(c) to provide a more intuitive picture of this trend.  All the $\it{\rho}_{xy}$ curves exhibit a negative slope which increases under pressure, as seen in Fig. 4(b). This suggests that the electrons are dominant carriers in the whole pressure range. The nonlinear field dependence of $\it{\rho}_{xy}$ shows an evident two-carrier transport. We use the two-band model to analyze the nonlinear Hall resistivity \cite{ali2014non,luo2015hall},
\begin{eqnarray}
\rho_{xx}&=&\frac{(n_e\mu_e+n_h\mu_h)+(n_e\mu_e\mu_h^2+n_h\mu_h\mu_e^2)B^2}{(n_e\mu_e+n_h\mu_h)^2+\mu_h^2\mu_e^2(n_h-n_e)^2B^2}\cdot\frac{1}{e}\\
\rho_{xy}&=&\frac{(n_h\mu_h^2-n_e\mu_e^2)+\mu_e^2\mu_h^2(n_h-n_e)B^2}{(n_e\mu_e+n_h\mu_h)^2+\mu_h^2\mu_e^2(n_h-n_e)^2B^2}\cdot\frac{B}{e}
\end{eqnarray}
where $n_h$ $(n_e)$ and $\mu_h$ $(\mu_e)$ are carrier density and mobility of hole (electron), respectively. The longitudinal resistivity $\it{\rho}_{xx}$ and Hall resistivity $\it{\rho_{xy}}$ are fitted simultaneously to derive the best self-consistent fitting parameters (see Fig. 4(d)). As pressure reaches to 5.3 GPa, the carrier density of electron reaches its maximum while the carrier density of hole exhibits a distinct kink. These anomalous behaviors correspond to the pressure region where ${\it T}_{\rm c}$ reaches to its first maximum. Upon further compression, both the carrier densities of electron and hole experience a rapid decrease. Until 15.0 GPa, the reduction starts to slow down and shows a tendency to gradually stablize. These observation suggests the change of electronic structure is involved as pressure increases.

The temperature-pressure ($\it{T}-\it{p}$) phase diagram is summarized in Fig. 5. Two superconducting domes can be clearly seen with SC-I under lower pressures and SC-II under higher pressures. In the SC-I phase, ${\it T}_{\rm c}$ exhibits a rapid increase and followed by a quick suppression. The superconducting transition becomes undetectable at $\sim$10 GPa. In the SC-II phase, superconductivity reemerges and then ${\it T}_{\rm c}$ gradually decreases until the superconducting transition is completely suppressed upon further compression. The superconductivity reemerges at $\sim$13 GPa, which is consistent with the pressure region that the magnetoresistence dramatically increases and carrier density becomes weakly pressure dependent. Therefore, the reemergent superconductivity is closely related to the evolution in the carrier density and, consequently, the electronic structure.

The observation of a superconducting dome under pressure has been extensively reported in unconventional superconductors \cite{norman2011the, kusmartseva2009pressure,cheng2015pressure,sun2016dome}. In comparison, the presence of two distinct superconducting domes is a relatively uncommon phenomenon. While the first superconducting dome normally associated with the suppression of some ordered phase, thus some kind of quantum fluctuation, the underlying mechanisms of the second superconducting dome remain yet to be fully elucidated. Typically, the second dome is accompanied by changes in spin configuration, band structure or structual phase transition. For instance, in heavy fermion superconductor CeCu$_2$(Si$_{1-x}$Ge$_x$)$_2$, the low-pressure superconducting dome occurs around an antiferromagnetism quantum critical point, suggesting magnetically mediated pairing, while the high-pressure superconducting dome straddles a weak first-order volume collapse and is believed to associated with the valence fluctuation of Ce ion \cite{yuan2003observation,yuan2006non,stockert2011magnetically,holmes2004signatures}. In the case of iron chalcogenides $\it{A}$$_x$Fe$_{2-y}$Se$_2$ [$\it{A}$ = K, (Rb, Tl)], after ${\it T}_{\rm c}$ drops from the first maximum of 32 K at 1 GPa, a second superconducting dome suddenly appears above 11.5 GPa with ${\it T}_{\rm c}$ reaching the maximum of 48 K \cite{sun2012reemerging}. X-ray emission spectroscopy and diffraction measurements display the change of Fermi surface topology with a possible tetragonal to collapsed tetragonal phase transition at the transition pressure \cite{yamamoto2016origin}.
As for the vanadium-based kagome superconductors $\it{A}$V$_3$Sb$_5$ ($\it{A}$ = Cs, Rb and K), the origin of the SC-I phase is strongly related to the CDW instability, and the SC-II phase is attributed to pressure-induced Lifshitz transition and enhanced electron-phonon coupling \cite{yu2021unusual,chen2021double,zheng2022emergent,Chen2021highly,yu2022pressure}.

\begin{figure}
	\includegraphics[width=8.6cm]{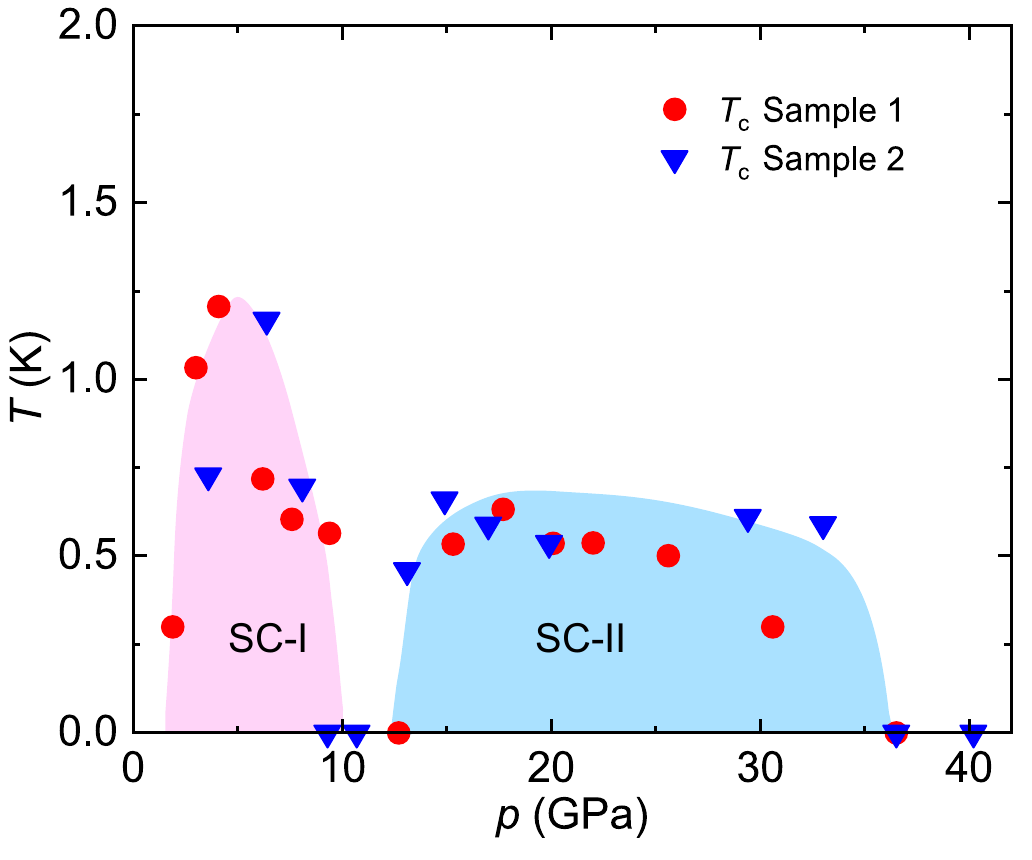}
	\caption{ Temperature-pressure phase diagram of CsTi$_3$Bi$_5$. Two superconducting domes can be clearly seen, with the SC-I phase under lower pressures and the SC-II phase under higher pressures.}
	\label{fig.5}
\end{figure}

The phase diagram of CsTi$_3$Bi$_5$ in Fig. 5 is quite similar to that of  CsV$_3$Sb$_5$, suggesting that these two kagome systems may share a common origin of double-dome superconductivity. While no CDW order is observed in CsTi$_3$Bi$_5$, recent STM measurements do reveal nematic quasiparticle interference patterns \cite{li2022electronic}. These patterns are dominated by the $\it{d}$ orbitals of Ti, suggesting a potential origin from the intriguing orbital bond nematic order. The emergence of such intra- and inter-orbital bond nematic orders can be ascribed to the extended Coulomb interaction in analogy to the nematic phase of FeSe \cite{jiang2016interatomic,zhang2015observation}. In this context, the first superconducting dome observed in CsTi$_3$Bi$_5$ may relate to the instability of the nematic order under pressure.

For CsV$_3$Sb$_5$, Yu $\textit{et al.}$ found that the emergence of the second superconducting dome under high pressure is inherently linked to the interlayer Sb2-Sb2 interactions \cite{yu2022pressure}. As the interlayer distance decreases, the interlayer interaction becomes more pronounced at high pressure and leads to the eventual bonding of the Sb2 atoms on the honeycomb structure \cite{yu2022pressure}. As a result, the $\it{p_z}$ orbital of Sb2 gradually passes through the Fermi level, resulting in the enhancement of the density of states and consequently the increase of ${\it T}_{\rm c}$ \cite{yu2022pressure}. Therefore, the abnormal evolution of superconductivity can be attributed to the formation of interlayer Sb2-Sb2 bonding that enhances the system's three dimensionality at high pressure \cite{yu2022pressure}. For CsTi$_3$Bi$_5$, recent calculations show the emergence of a special valley and dome in the ${\it T}_{\rm c}$ under pressure, which is due to the crossover from quasi-two-dimensional to three-dimensional isotropic compression within the range of 10-20 GPa \cite{yi2023superconducting}. In Fig. 4(d), the clear deviation of carrier densities above 15 GPa from the low-pressure declining trend is likely correlate to this dimensional crossover and the emergence of the second superconducting dome.

In summary, we investigate the electronic transport properties of kagome metal CsTi$_3$Bi$_5$ under pressures up to 40 GPa. While CsTi$_3$Bi$_5$ displays metallic behavior under ambient pressure, superconductivity is induced by pressure and ${\it T}_{\rm c}$ displays remarkable double domes in the phase diagram. Such a double-dome superconductivity under pressure is very similar to that in $\it{A}$V$_3$Sb$_5$ ($\it{A}$ = Cs, Rb and K), revealing that it is a universal phenomenon in these kagome metals. For the underlying physics of the two superconducting domes, the first one is strongly related to some competing order (nematic order in CsTi$_3$Bi$_5$, likely), and the second one may be due to the variation of electronic band structure caused by dimensional crossover under high pressure.

This work was supported by the Natural Science Foundation of China (Grant No. 12174064) and the Shanghai Municipal Science and Technology Major Project (Grant No. 2019SHZDZX01). Y.F.G. was supported by the open project of Beijing National Laboratory for Condensed Matter Physics (Grant No. ZBJ2106110017) and the Double First-Class Initiative Fund of ShanghaiTech University.

J. Y. Nie, X. F. Yang, X. Zhang and X. Q. Liu contributed equally to this work.

\noindent $^\#$ E-mail: yangxiaofan$@$fudan.edu.cn\\
\noindent $^\S$ E-mail: xuzhang$\_$fd$@$fudan.edu.cn\\
\noindent $^\dag$ E-mail: guoyf$@$shanghaitech.edu.cn\\
\noindent $^\ddag$ E-mail: shiyan$\_$li$@$fudan.edu.cn\\

\end{document}